\documentstyle[aps,prl,preprint]{revtex}
\begin{document}
\draft
\title{Manipulation of the Spin Memory of Electrons in n-GaAs}
\author{R. I. Dzhioev, V. L. Korenev\cite{Ioffe}, I.A.Merkulov, B. P. Zakharchenya}
\address{A. F. Ioffe Physical Technical Institute, St. Petersburg, 194021 Russia}
\author{D. Gammon, Al.L. Efros, D.S. Katzer}
\address{Naval Research Laboratory, Washington DC 20375}
\date{\today}
\maketitle

\begin{abstract}
We report on the optical manipulation of the electron spin relaxation time
in a GaAs based heterostructure. Experimental and theoretical study shows
that the average electron spin relaxes through hyperfine interaction with
the lattice nuclei, and that the rate can be controlled by the
electron-electron interactions. This time has been changed from 300 ns down
to 5 ns by variation of the laser frequency. This modification originates in
the optically induced depletion of  n-GaAs layer.
\end{abstract}

\pacs{PACS: 71.35.+z, 73.61.Ey, 78.55.Cr}

\narrowtext

 A long spin relaxation time for electrons, $\tau _{s}$, \cite{1,2,3} could hold
promise for magnetoelectronics and quantum information applications \cite
{4,5}. This time can be controlled by an external magnetic field \cite{6}
, by choice of heterostructure geometry \cite{7}, by doping level 
\cite{2,6,8}, and by a gate voltage that changes the electron concentration
in the two-dimensional channel \cite{9}.

This letter is devoted to the {\it optical} manipulation of the spin memory
of electrons. We show, using optical orientation methods, that the spin
relaxation rate in a GaAs-based heterostructure can be modified by a factor
of 100 by variation of the laser frequency. The comparison between experiment
and theory shows that the observed effect is the result of turning on or off
of the hyperfine channel of spin relaxation of electrons localized on
shallow donors. The giant changes of spin relaxation time allow one to
extend the concept of spin dynamics engineering in quantum wells \cite{9}
to  quasibulk semiconductors having much longer spin lifetimes.

The optical orientation experiments were carried out on a sample with a
semi-insulating GaAs substrate, on which a 500\,nm thick GaAs buffer and a 
25\,nm thick AlAs barrier separating the substrate from the main structure
were MBE-grown. Following this is a 100 nm thick GaAs layer capped by a 
25\,nm  thick Al$_{0.3}$Ga$_{0.7}$As barrier, above which a series of five
quantum wells of different thickness were prepared (Fig.\ref{f1}b). The
sample, though nominally undoped, has a background doping level of order 
$10^{14}$\,cm$^{-3}$ n-type from Hall measurements on test structures. The
dominant photoluminescence (PL) at the wavelength close to the GaAs band
edge is due to the 100\,nm layer. The sample was placed in a liquid-helium
cryostat and pumped by a tunable Ti-sapphire laser, with the circular
polarization of light being alternated in sign at a frequency of 26.61\,kHz 
 with a photoelastic quartz modulator. The geomagnetic field was
compensated to less than 0.1\,G at the sample. The PL
polarization was measured in the reflection geometry with a fixed
circular-polarization analyzer by a double-grating
spectrometer. The electronics provided
measurement of the {\it effective} degree of circular polarization $\rho
=\left( I_{+}^{+}-I_{+}^{-}\right) /\left( I_{+}^{+}+I_{+}^{-}\right) $,
where $I_{+}^{+},I_{+}^{-}$ $\,$are the intensities of the $\sigma ^{+}$ PL
component under $\sigma ^{+}$ and $\sigma ^{-}$ pumping, respectively.

When the excitation energy, $h\nu $ , is below the absorption edge of the quantum
wells, the PL spectrum of the 100 nm GaAs layer (curve $1$ in Fig.1a at 
$h\nu =1.520$\,eV) is dominated by two lines, corresponding to the
recombination of the free exciton (X) and of the exciton bound to a
neutral donor (D$^{0}$X) \cite{11}. The GaAs PL spectrum changes
drastically when the laser frequency is in resonance with the heavy-hole
exciton transition (HHX) of the quantum well (14\,nm) nearest to the 
100\,nm layer (curve 2 at $h\nu =1.534$\,eV, Fig.1a). The intensity
of the exciton line becomes 3 times larger whereas the intensity of the 
D$^{0}$X complex decreases. The same changes take place when one excites the
light-hole (LHX) exciton ($h\nu =1.540$\,eV). Curve 3 shows the PL
excitation spectrum (PLE) of the GaAs layer exciton line when the laser
frequency is tuned through the exciton resonances in the 14 nm QW.

One can see sharp enhancements in the PL intensity of the 100\,nm
GaAs exciton peak under resonant excitation of heavy-hole and
light-hole excitons in the QW. The PLE spectrum of the D$^{0}$X
complex evolves with phase opposite to that of the exciton PLE. There is 
also a striking interplay between the exciton and trion lines in
the spectrum of the QW PL. The solid line in Fig.3c shows that
the PL trion peak (1.5326\,eV) dominates over the HHX PL (1.534\,eV)
when one excites the LHX transition in the 14\,nm QW. The full (open)
circles show the PLE spectrum of the QW exciton (trion) peak. One can see
that resonant excitation of excitonic transitions favors the trion line
whereas the exciton line is stronger at excitation out of resonance. The quenching
of D$^{0}$X in the PL of bulk GaAs and the simultaneous sharp enhancement of
the trion in the PL spectrum of the QW indicate that the GaAs layer is
depleted of electrons whereas the 14\,nm QW is enriched of them \cite{14}.
 This phenomenon disappears when the pump density is below 0.1\,mW/cm$^{2}$.
 In this case the PL of the GaAs layer consists of D$^{0}$X and X
lines and the QW PL reveals only the exciton line. These results can be
understood qualitatively with the use of Fig.1b. Without excitation of the
QW, the GaAs layer is charged negatively: besides donor-bound electrons
there are extra free electrons escaping the donors from the surrounding barriers
and wells. The excitation of the QW induces their redistribution, depleting
the GaAs layer and filling the QW. One possibility is the tunneling of the
QW photoexcited holes into the 100 nm GaAs layer and recombination with electrons.
Another possibility is the annihilation of one QW exciton with the
simultaneous ionization of another one through Auger-process. In this case the hole goes
into the GaAs by jumping over the barrier while the electron goes back due to
repulsion from the negatively charged layer.  

Overall, the experimental data
clearly show that the density of electrons in the 100 nm layer is reduced and that in the QW is 
increased  independent of the specific mechanism responsible for the recharging effect. 
In this paper we will {\it exploit} this
phenomenon to manipulate the 100\,nm layer electron concentration
optically in order to trace the electron {\it spin} 
relaxation in the 100\,nm GaAs layer. We show that the depletion of 
the 100\,nm GaAs
switches on a fast electron spin relaxation mechanism (5\,ns) involving
lattice nuclei, whereas the presence of extra electrons in the layer
strongly suppresses it (up to 300\,ns).

The electronic spin relaxation in GaAs can be probed accurately with the use
of the Hanle effect under optical pumping conditions \cite{6}.
Circularly polarized light creates electronic spin polarization oriented
along the exciting beam. The degree of circular polarization, $\rho $, is
determined by the projection, $S_{z},$ of the mean electron spin on the pump
beam direction: $\rho =S_{z}$. An external magnetic field in Voigt geometry
(i.e. a transverse field) decreases the steady state PL polarization due to
Larmor precession with frequency $\omega =\mu _{B}g_{e}B/\hbar$ ($\mu
_{B} $ is the Bohr magneton, $g_{e}=-0.44$ \cite{15} is the electron $g$
-factor). The field dependence of $\rho $ is described by a Lorenzian curve
with halfwidth, $B_{1/2}=\hbar /\mu _{B}g_{e}T_{s}$ (Hanle effect), where
the electron spin lifetime $T_{s}=\left( 1/\tau +1/\tau _{s}\right) ^{-1}$
is determined by recombination $1/\tau $ and spin relaxation $1/\tau _{s}$
rates. The electron lifetime in n-type semiconductors depends strongly on
the pumping intensity (photo-hole concentration), and in the low pumping
limit is very long $\left( \tau >>\tau _{s}\right) $. In this limit the spin
lifetime coincides with the spin relaxation time, e.g. $T_{s}\left(
W=0\right) =\tau _{s}$, and determines directly the linewidth of the Hanle
curve.

Figure 2 shows the Hanle depolarization curves measured at different laser
energies with fixed power density, $W=40$\,mW/cm$^{2}$. Curve 1 shows the data
measured under quasiresonant creation of the exciton of the 100\,nm layer. 
The halfwidth value, $B_{1/2}=2.9$\,G, corresponds to $
T_{s}=90$\,ns. The spin lifetime becomes shorter when one excites above the
GaAs energy gap (curve 2): $T_{s}=40$\,ns ($B_{1/2}=6.8$\,G). The insert in Fig.2 shows the
dependence of $1/T_{s}$ on $W$ under quasiresonant
excitation of the GaAs exciton. The low power limit of
this dependence gives $\tau _{s}=\left( 290\pm 30\right)$\,ns.

Increasing the laser frequency, with power fixed, up to the heavy-hole exciton
energy in the 14\,nm QW brings no drastic changes to the halfwidth of the Hanle curve, $B_{1/2}$. The
situation changes dramatically when either the QW heavy-hole or light-hole
exciton is excited: $B_{1/2}$ broadens to 50\,G. As an
example we present the data  obtained under resonant
LHX excitation (curve 3 in Fig.2). When the exciting frequency is between
the HHX and LHX, the QW is not excited and the polarization behavior of
the $100$\,nm layer luminescence is similar to curve 2. Further
increasing the laser frequency above the QW resonances increases the QW
excitation and  $B_{1/2}$ again approaches to $50$\,G. We
conclude that there is a large quantitative difference in $B_{1/2}$
 of the 100\,nm layer when the QW is excited (the layer is
depleted of electrons) or not (extra free electrons fill up the layer). The change of $B_{1/2}$ 
can not be connected with spin transfer from the QW across the 25\,nm AlGaAs barrier
during the recharging process.
Direct evidence against this is that both the zero-field polarization
values ($+2\,\%$) and the Hanle effects in the GaAs layer are identical
under LHX (Fig.2, curve 3) and HHX (Fig.3, b) excitation of the
14\,nm QW. However the QW exciton polarization has opposite signs in
these two cases due to selection rules (experimental values are $-10\,\%$
for LHX and $+20\,\%$ for HHX excitations for positive laser
polarization).

Remarkably, {\it qualitative} differences between the two regimes appear
when the magnetic field is applied in the Faraday geometry, i.e. under a
longitudinal field. In Figure 3, polarization data taken in the Faraday and
Voigt geometry are compared for the two regimes: when the QW is not excited
and extra electrons remain in the 100 nm layer (Fig.3a); and when the QW is
excited and the electron population is reduced (Fig.3b). With the extra
electron population, there is practically no magnetic field dependence in
the Faraday geometry (upper part of Fig.3a). In contrast, when the electron
population is reduced, application of the magnetic field in the Faraday
geometry increases the polarization by a factor of 2.5 (upper part of
Fig.3b). Moreover, although there is no correlation in the change in
polarization in the two geometries in Fig. 3a, the characteristic transverse
and longitudinal magnetic field values that decrease and increase,
respectively, the electron polarization are surprisingly similar in Fig. 3b.
This difference in behavior reflects a qualitative difference in the spin
relaxation process in the two experimental regimes, i.e. with high (Fig.3a)
and low (Fig.3b) electron concentrations in the GaAs layer.

The experimental data shown in Fig.3 correspond to the two limiting
cases of the electron spin relaxation in a rapidly fluctuating or static
random magnetic field. The spin relaxation can be viewed as the
result of the action of random local magnetic fields on electron spin 
\cite{16} and depends both on the amplitude of the random field and on its
fluctuation rate. In the  static limit the characteristic time ($\tau _{c}$) of the local
field fluctuation is much longer than the period of electron spin precession
in this field ($\omega _{f}^{-1}$) and the electron spins undergo precession in
the static fields distributed randomly. 
 The average spin of the ensemble relaxes to 1/3 of its initial value
with a characteristic time $\tau _{s}\approx \omega _{f}^{-1}$ \cite{17}. 
The longitudinal external field eliminates spin relaxation, and the transverse field 
 depolarizes electrons when it overwhelms the local field,
i.e. $\omega \geq \omega _{f}$. In the opposite limit: $\tau _{c}<<\omega _{f}^{-1}$ (the so-called motional
narrowing case \cite{18}), dynamic
averaging takes place and the spin relaxation slows down: $\tau
_{s}^{-1}\approx \omega _{f}^{2}\tau _{c}$. In this case, the transverse
magnetic field that depolarizes the electron spin (Hanle
effect) satisfies the condition $\omega \geq \omega _{t}\equiv \tau
_{s}^{-1}\approx \omega _{f}^{2}\tau _{c}$. The longitudinal field 
required to suppress spin relaxation satisfies the condition: $\omega \geq \omega _{{\it l}}\equiv\tau
_{c}^{-1}$,(see  Ref.\cite{19}). It
happens in much greater magnetic fields because $\omega _{t}=\tau
_{s}^{-1}\approx \omega _{f}^{2}\tau _{c}=\left( \omega _{f}\tau _{c}\right)
^{2}\omega _{{\it l}}<<\omega _{{\it l}}$. Therefore, the experimental
result of Fig.3a, with additional electrons in the layer, corresponds to
the motional narrowing case $\left( \omega _{{\it l}}>>\omega _{t}\right) $,
and the result of Fig.3b, with electrons removed, corresponds to the
static limit $\left( \omega _{{\it l}}\approx \omega _{t}\right) $.

The experimental data obtained in the static case enables us to deduce that
the main spin relaxation mechanism of electrons in this experiment is the
hyperfine interaction of localized electrons with lattice nuclei. 
 It follows from Fig.3b that the characteristic precession frequency
of the electron spins in the random local field is $\omega _{f}\approx
2\cdot 10^{8}\,s^{-1}$. Therefore $\tau _{c}>\tau _{s}\approx \omega
_{f}^{-1}\approx 5$\,ns. None of the known spin relaxation mechanisms of
free carriers and excitons (Chap.3 in Ref.\cite{6}) can have such long
correlation times. However, bound electrons at low electron concentration
and low temperature may spend a long time at a given donor. Moreover, the $
\omega _{f}$ value agrees well with the calculated precession frequency of
spins of the donor-bound electrons in the random hyperfine fields of
surrounding nuclei as we now show.

To describe the average electron spin polarization in the low concentration
limit we consider an evolution of the electron spin, ${\bf S}_{n}$,
localized at the $n$-th donor. It is described by the Bloch equation
\cite{6} with the precession frequency,{\bf \ }${\bf \Omega }_{n}={\bf 
\omega }+{\bf \omega }_{fn}$:
\begin{equation}
\frac{d{\bf S}_{n}}{dt}=\frac{{\bf S}_{i}-{\bf S}_{n}}{\tau _{c}}+{\bf 
\Omega }_{n}\times {\bf S}_{n}=0  \label{e1}
\end{equation}
where ${\bf S}_{i}$ is the initial electron spin value in the moment of
trapping. During the correlation time, $\tau _{c}$, the electron spin
dynamics is determined by the external and the local nuclear magnetic fields
only. The static nuclear fields are described by a Gaussian distribution, $
\Phi \left( B_{N}\right) =\pi ^{-3/2}\Delta ^{-3}\exp \left(
-B_{N}^{2}/\Delta ^{2}\right) $, with a dispersion, $\Delta =54$\,G, given
by the root-mean-square of the random nuclear field \cite{16}. The average nuclear 
polarization is zero because we are modulating the polarization of exciting light.
Averaging
the solution of Eq.\ref{e1} over the nuclear magnetic field distribution 
\cite{17} results in a theoretical dependence of circular polarization on
the longitudinal and transverse magnetic field that are presented in Fig.3b
by dashed and solid lines, respectively. The {\it only} fitting parameter
for the curves in Fig.3b is $\tau _{c}=17$\,ns, which satisfies the static
condition, $\tau _{c}>\omega _{f}^{-1}$. This value is determined by the
processes of donor electron ionization, recombination, exchange scattering
by free electrons, etc. The good agreement between experiment and theory
confirms that the interaction with nuclei is the main mechanism of the
electron spin relaxation for the localized carriers.

The much longer spin relaxation time in the motional narrowing case ($\tau
_{s}\approx 300$\,ns) arises from the increase in electron concentration.
Recall that the presence of extra free electrons in addition to the
``intrinsic'' donor-bound ones, comes from donors in the barriers and
quantum wells surrounding the GaAs layer. The suppression of spin
relaxation is caused by the decrease of the correlation time (from $\tau
_{c}>$5\,ns to $\tau _{c}\approx 0.1$\,ns \cite{21}) and can be
explained by scattering processes between the free and bound electrons
discussed in Ref.\cite{22}. These processes average out the inhomogeneous
nuclear field and suppress spin relaxation by two orders of magnitude. The
exciton PL serves as a detector of spin polarization of donor-bound
electrons \cite{23}.

In conclusion, we have shown that the spin relaxation time of donor-bound
electrons in a thick GaAs layer depends strongly on the concentration of
extra electrons. This enabled us to manipulate the spin relaxation rate by
two orders of magnitude. The presence of extra free electrons in the
conduction band suppresses strongly the spin relaxation of the donor bound
electrons because the inhomogeneous nuclear field is averaged out, leading
to an extremely long spin relaxation time of $0.3\,\mu$s.  Decrease of the electron concentration
breaks up the electron spin system into an ensemble of individual
donor-bound electrons, and the electronic spin relaxation is the result of
dephasing by the randomly distributed hyperfine fields. In our experiments,
the average of the nuclear spin polarization is zero but its dispersion is
approximately 54\,G, corresponding to an inhomogeneous dephasing rate of 
5\,ns. The effect of inhomogeneous dephasing can be eliminated with the
use of the spin echo technique Ref.\cite{18}. 
 This case may be important in efforts to implement quantum
computing using localized electronic spins, which require not only long
coherence times but also must be localized in order to be individually
addressed.

Partial support of the Russian Basic Research Foundation (grants 00-02-16991
and 01-02-17906), CRDF (grant RP1-2252), DARPA/SPINS, and ONR is gratefully
acknowledged.

\begin{figure}[tbp]
\caption{(a): PL and PLE spectra of the 100 nm thick GaAs layer. PL curve 1
was measured at excitation, $h\nu=1.520$\,eV (only GaAs layer was excited),
and curve 2 was measured at $h\nu=1.534$\,eV (the heavy-hole exciton in the
14 nm QW was excited resonantly). PLE curve 3 was measured at the exciton
(X) energy (1.5155 eV). Maxima in the PLE spectrum correspond to the
resonant excitation of the heavy-hole (HHX) and light-hole (LHX) excitons in
the 14 nm QW. (b) The part of the structure containing the 100 nm thick GaAs
layer and the 14 nm QW. Without illumination, additional electrons from
barriers and quantum wells are accumulated in the GaAs. (c) PL and PLE
spectra of the 14 nm thick QW. PL (solid curve) was measured at excitation, 
$h\nu=1.540$\,eV (LH-exciton). PLE of the exciton line (1.534 eV, full
circles) and trion line (1.5326 eV, open circles).}
\label{f1}
\end{figure}

\begin{figure}[tbp]
\caption{The Hanle effect of the 100 nm GaAs electrons at three excitation
conditions: (1) with quasi-resonant excitation of GaAs excitons; (2) with
excitation above the GaAs energy gap but in the transparency region of QWs;
(3) with resonant excitation of the LH-exciton in the 14 nm QW. Insert:
power dependence of the reciprocal spin lifetime.}
\label{f2}
\end{figure}

\begin{figure}[tbp]
\caption{Magnetic field dependence of the circular polarization in Voigt
(full circles) and Faraday (open circles) geometries under two excitation
conditions at T=4.2 K. Polarization degree is normalized to
the zero field values, $\rho(0)$, of the exciton PL (1.5155 eV) in GaAs. (a)
GaAs layer is excited only ($\rho(0)$=15\% ). The solid curve is a
Lorentzian fit with $B_{1/2}=4.2$\,G. The dashed line is a guide to the eye.
(b) Resonant excitation of the HH-exciton in the 14 nm QW ($\rho(0)$=2\% ).
The dashed and solid lines are the theoretical dependences of the electron
spin calculated from Eq. (1) with the only fitting parameter, 
$\tau _{c}$=17\,ns.}
\label{f3}
\end{figure}


\begin{references}
\bibitem{19}  Here we do not take into account the effect of magnetic field
on spin relaxation due to cyclotron motion of the electrons (Chap.3 in Ref.
\cite{6}). It is necessary to assume an unlikely long momentum scattering
time, $\tau _{p}\approx 70$\,ps, for the effect to be essential at the field 
50\,G.
\end{references}
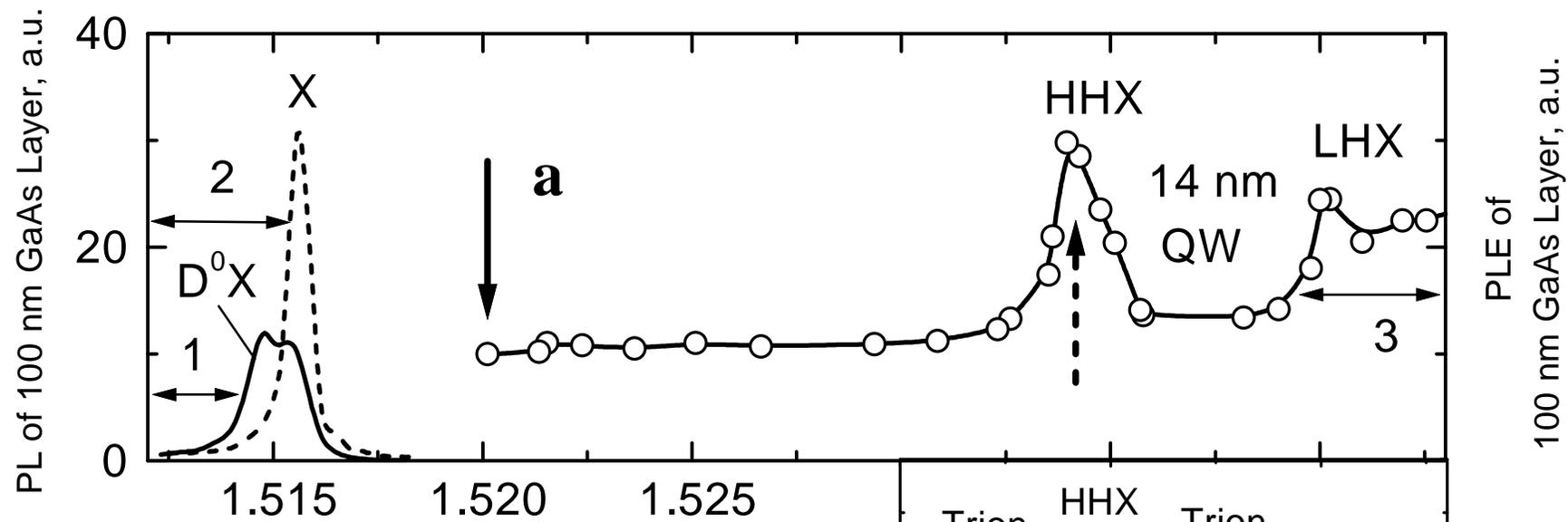
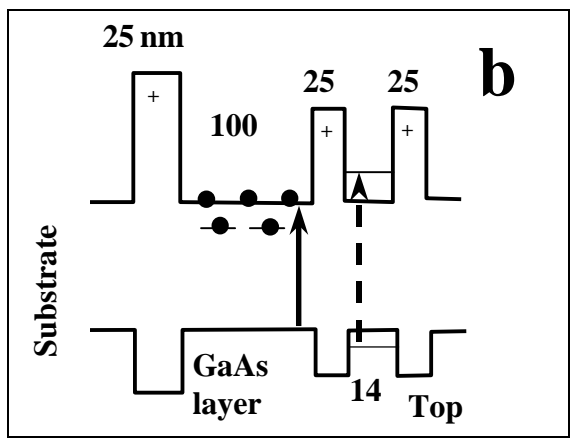
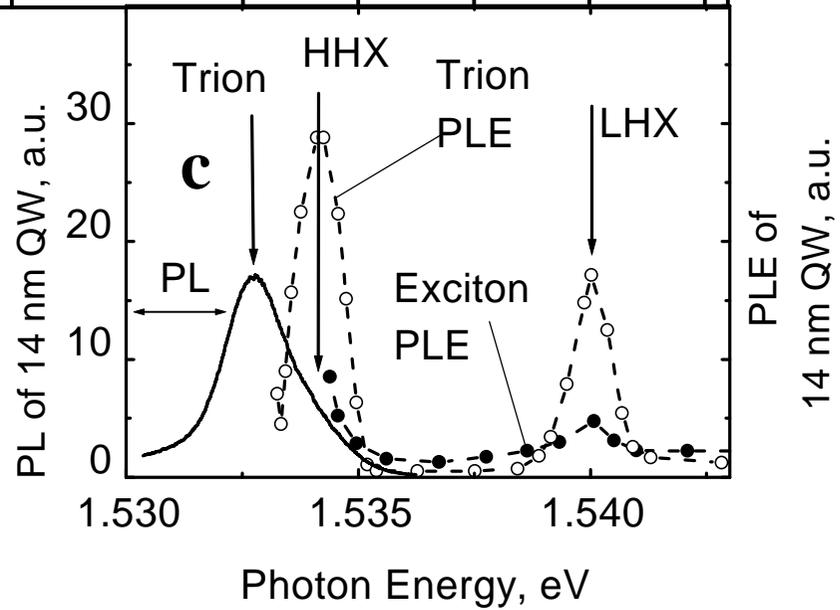
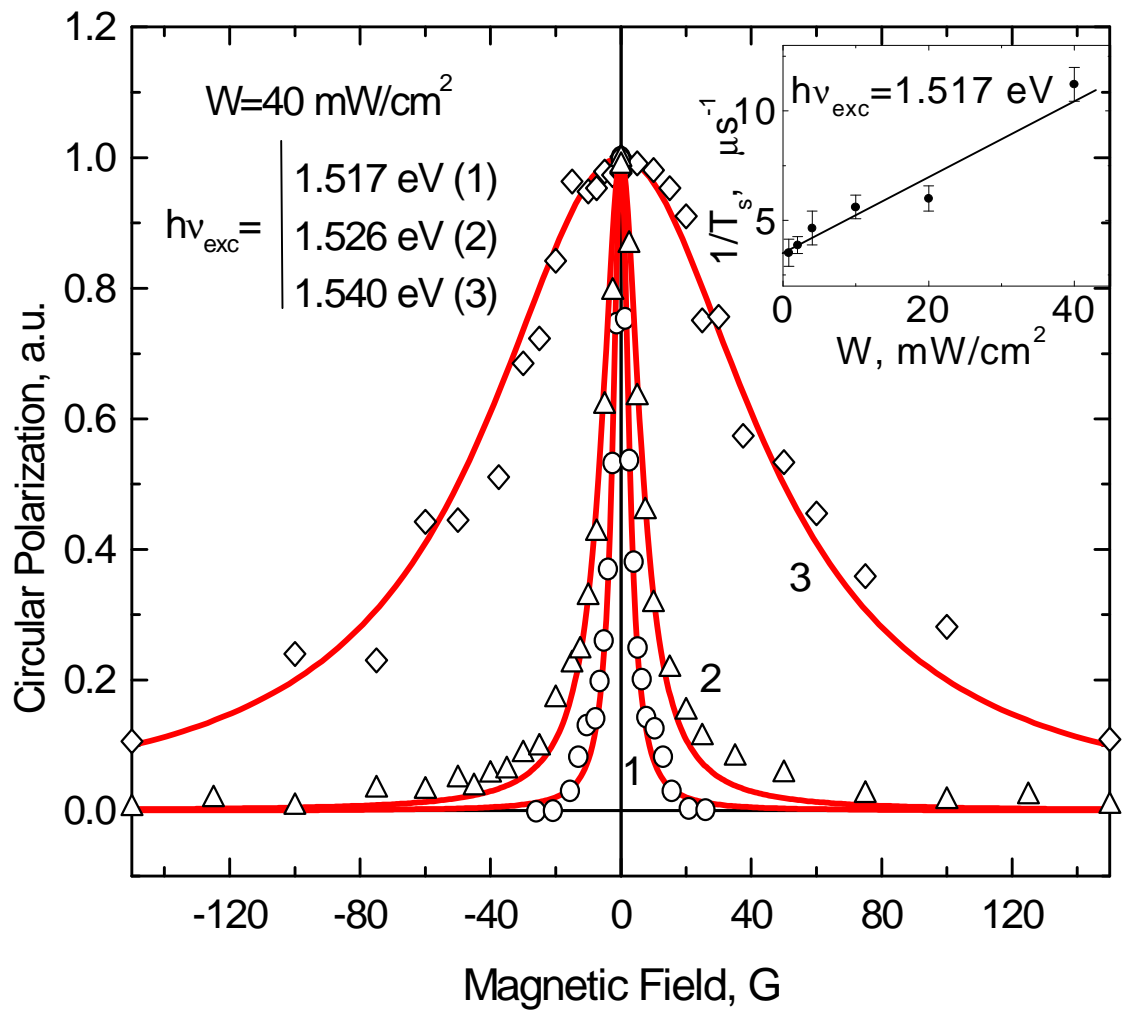
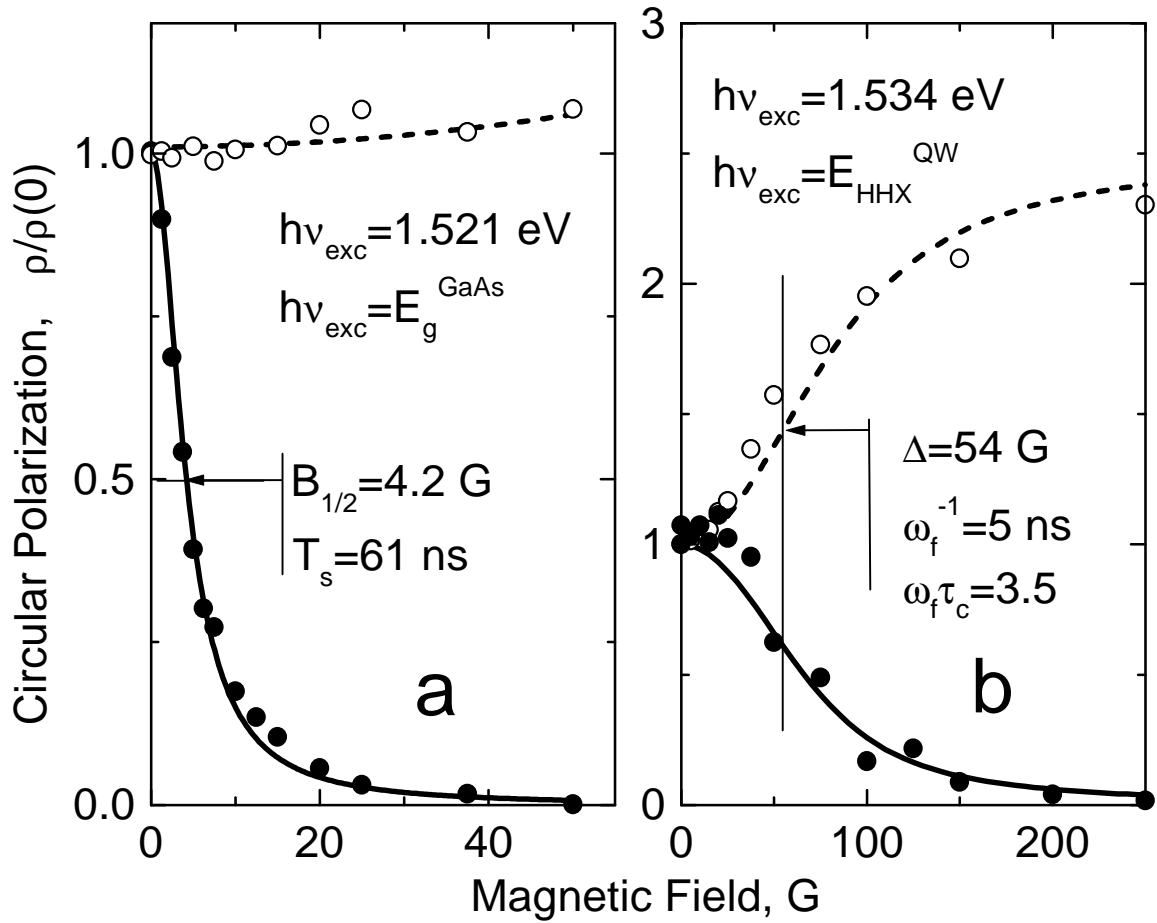
\end{document}